\documentclass[aps,reprint,twocolumn,amsmath,amssymb,showpacs,longbibliography]{revtex4-1}
\usepackage{graphicx,bm} \usepackage{color} 
\usepackage{ragged2e}
\usepackage{multirow}
\usepackage{mathrsfs}

\begin{document}

\title{Topological edge states of massless fermion with non-quantized and zero Berry phases}
\author{F. R. Pratama$^{1,2}$}
\email{pratama.fr@aist.go.jp}
\author{Takeshi Nakanishi$^{1}$}
\email{t.nakanishi@aist.go.jp}

\affiliation{$^{1}$Mathematics for Advanced Materials-OIL, AIST, 2-1-1 Katahira, Aoba, 980-8577 Sendai, Japan\\
$^{2}$Advanced Institute for Materials Research, Tohoku University, 2-1-1 Katahira, Aoba, 980-8577 Sendai, Japan }

\begin{abstract}

We investigate the bulk-boundary correspondence for massless Dirac fermion in $\alpha$-${T}_3$ lattice where the Berry phase can be continuously tuned from $\pi $ (graphene) to $0$ ($T_3$ or dice lattice) without modifying the energy dispersion. The topological origin of edge states is revealed by unitary transform of the Hamiltonian, which maps $\alpha$-${T}_3$ zigzag ribbon (ZR) into stub Su-Schrieffer-Heeger (SSH) chain. In the Majorana representation of the eigenstates, the $\mathbb{Z}_2$ invariant is manifested by azimuthal winding numbers on the Bloch sphere. Particularly, $T_3$ ZR exemplifies a topologically nontrivial system with zero Berry phase. Contrary to the conventional topological insulators, band gap closing in the corresponding stub SSH chain is not accompanied by a topological phase transition. 

\end{abstract}
\date{\today}
\maketitle

Topological matters host robust edge states~\cite{hasan2010} that are promising for applications in quantum computing and efficient electronic devices~\cite{luo2022topological}. In the Chern insulators~\cite{thouless1982,haldane1988model}, time-reversal symmetry (TRS) is broken and the number of edge states is given by $\mathbb{Z}$ invariants. In systems with preserved TRS, $\mathbb{Z}_2$ invariants~\cite{kane2005,bernevig2006,fu2011topological} differentiate between topological insulators (TIs) and trivial insulators. The simplest model for TI is the Su-Schrieffer-Heeger (SSH) chain of polyacetylene~\cite{su1979}, in which the ratio between intracell and intercell hopping integrals determine the Berry phase $\Gamma$ in one dimensional (1D) Brillouin zone (BZ), also known as the Zak phase~\cite{zak1989}. Here, the presence and absence of edge states are indicated by $\Gamma=\pi$ and $0$, respectively~\cite{delplace2011}. The existence of edge states in graphene zigzag ribbon (ZR) can also be inferred from the Zak phase~\cite{delplace2011}. By fixing a proper gauge for the Hamiltonian, graphene is mapped into the SSH chain~\cite{ryu2002,bellec2014manipulation,st2021measuring,gong2023mapping}. The mapping establishes a relation between the conserved momentum of graphene ZR and the hopping parameter of the SSH chain, and therefore, the system is reduced from 2D to 1D.



\begin{figure}[t]
\begin{center}
\includegraphics[width=83mm]{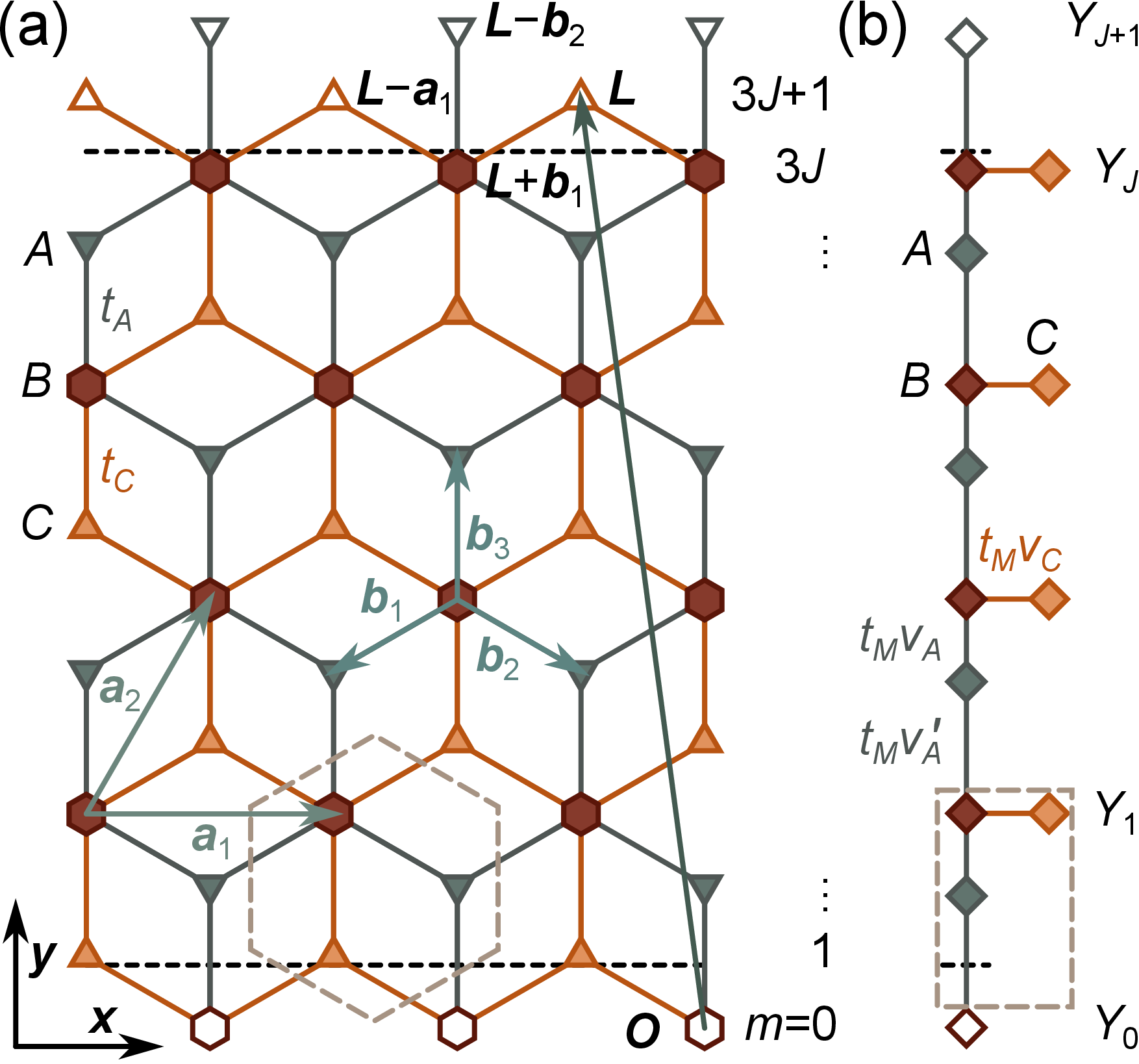}
\caption{ (a) The $\alpha$-$T_3$ ZR. The hopping parameter between the $B$ and $A$ ($C$) sites is $t_A$ ($t_C$), where $t_C/t_A=\alpha\in[0,1]$. The dashed hexagon depicts the unit cell. $\boldsymbol{b}_1=-a(1/2, 1/2\sqrt{3})$, $\boldsymbol{b}_2=a(1/2,-1/2\sqrt{3})$, and $\boldsymbol{b}_3=a(0,1/\sqrt{3})$ are the nearest-neighbor vectors. $\boldsymbol{a}_1=a(1,0)$ and $\boldsymbol{a}_2=a(1/2,\sqrt{3}/2)$ are the primitive vectors. $m=1,2,\dots, 3J$ indicate the position of site in the $y$ axis, where $J$ is the number of trimers. $\boldsymbol{L}=(L_x,L_y)$ connects the missing $B$ and $C$ sites at $m=0$ and $m=3J+1$, respectively, where $L_y=\sqrt{3}(J+1)a/2-a/\sqrt{3}$. (b) The stub SSH chain. $t_Mv_A$ ($t_Mv_A^\prime$) is the intracell (intercell) hopping parameters between the $B$ and $A$ sites. The hopping parameter between the $B$ and $C$ sites is $t_Mv_C$. The position unit cell (dashed rectangle) is $Y_j=j\sqrt{3}a/2$. }
\label{fig:lett-lattices}
\end{center}
\end{figure}

Consider a honeycomb lattice comprising $A$ and $B$ sites with hopping parameter $t_A>0$. The $\alpha$-$T_3$ lattice~\cite{raoux2014dia} is constructed by connecting the $B$ sites to the additional $C$ sites at the center of each hexagon with hopping parameter $t_C=\alpha t_A$ for $\alpha\in [0,1]$, as depicted in Fig. ~\ref{fig:lett-lattices}(a). Here, $\Gamma$ varies continuously between $\pi$ (graphene, pseudospin $S=1/2$) to $0$ ($T_3$ lattice, $S=1$) for $\alpha=0$ and $1$, respectively, without altering the energy bands when the energy is scaled by $t=\sqrt{{t_A}^2 + {t_C}^2}$. A question that naturally arises from this situation is whether the topological edge states remain without the quantized $\Gamma$. However, the bulk-boundary correspondence is yet to be formulated, even though the $\alpha$-$T_3$ lattice has been extensively studied due to its unconventional properties, e.g. tunable orbital susceptibilities from dia- to paramagnetism~\cite{raoux2014dia}, enhanced Klein tunneling~\cite{urban2011barrier,fang2016klein,illes2017klein}, and large Chern number $|\mathscr{C}|=2$ when the TRS is broken~\cite{wang2011nearly,dey2019floquet,dey2020unconventional, mohanta2023majorana}. Although the $\alpha$-$T_3$ and $T_3$ models were originally proposed~\cite{raoux2014dia,bercioux2009massless} to be realized in optical lattices, the $\alpha$-$T_3$ Hamiltonian can be employed to describe the optical properties of 2D $\mathrm{Hg}_{1-x}\mathrm{Cd}_x\mathrm{Te}$~\cite{orlita2014observation,malcolm2015magneto} at the critical doping $x\approx 0.17$. Furthermore, the $T_3$ lattice is predicted to occur in perovskite-based heterostructures\cite{wang2011nearly,koksal2023high} and strained blue-phosphorene oxide~\cite{zhu2016blue}.

In this study, an example of $\alpha$-$T_3$ ZR~[Fig. \ref{fig:lett-lattices}(a)] is mapped into stub SSH chain~\cite{bartlett2021illuminating,caceres2022experimental} [Fig. \ref{fig:lett-lattices}(b)] by unitary transform of the Hamiltonian. The stub SSH chain is constructed by connecting an additional site to one of the sites in the SSH dimer. In the Majorana representation of the eigenstates~\cite{majorana1932atomi,hannay1998berry,liu2014representation}, the $\mathbb{Z}_2$ invariant is defined from the azimuthal winding number on the Bloch sphere~\cite{bartlett2021illuminating}. Thus, the origin of edge states in the $\alpha$-$T_3$ ZR can be understood from the topological phase of the stub SSH chain. In particular, $T_3$ ZR is topologically nontrivial despite $\Gamma=0$. In contrast to the conventional TIs, the corresponding stub SSH chain is not subject to a topological phase transition by band gap closing. 





The $\alpha$-$T_3$ ZR in Fig. \ref{fig:lett-lattices}(a) is constructed by periodically translating the unit cell (dashed hexagon) along the $x$ direction with primitive vector $\boldsymbol{a}_1$. The width of the ZR is specified by $J$, which is the number of trimer along the $y$ axis. The vector $\boldsymbol{L}$ connects the missing $B$ and $C$ sites at $m=0$ and $3J+1$, respectively. The Hamiltonian in the momentum space $\boldsymbol{k}=(k_x,k_y)$ is given by
\begin{align}
h_{\boldsymbol{k}} = -t \begin{pmatrix}
     0   & \cos\vartheta f_{\boldsymbol{k}}   & 0\\
     \cos\vartheta f_{\boldsymbol{k}}^* & 0 & \sin\vartheta f_{\boldsymbol{k}}\\
     0 & \sin\vartheta f_{\boldsymbol{k}}^* & 0
     \label{eq:H2}
\end{pmatrix},
\end{align}
where $\vartheta = \tan^{-1}\alpha$, and $f_{\boldsymbol{k}}=\sum_{\ell=1}^{3}e^{-i\boldsymbol{k}\cdot\boldsymbol{b}_\ell}=|f_{\boldsymbol{k}}|e^{-i\varphi_{\boldsymbol{k}}}$. $\boldsymbol{b}_\ell$'s are the nearest-neighbor vectors. The eigenvalues of $h_{\boldsymbol{k}}$ include the energy bands of graphene $\varepsilon_{s,\boldsymbol{k}}=st|f_{\boldsymbol{k}}|$, where $s=+1$ ($-1$) corresponds to valence (conduction) band, and a flat band at zero energy $\varepsilon_{0,\boldsymbol{k}}=0$. 



By using the eigenstates $|\psi_{\boldsymbol{k},c}\rangle$ for the band $c\in\{s,0\}$, the Berry phase $\Gamma_c$ is calculated by integration along a path enclosing a Dirac point. The results are $\Gamma_{s,\tau}=\pi\tau\cos(2\vartheta)$ and $\Gamma_{0,\tau}=-2\pi\tau\cos(2\vartheta)$, where $\tau=+1$ ($-1$) for the $K$ ($K^\prime$) valley~\cite{raoux2014dia}. Thus, $\Gamma_c$'s are not quantized and differ for each valley when $\alpha\notin\{0,1\}$. 


Since there are several ways to define the unit cell (each consists of the unique configuration of trimer), distinct $\alpha$-$T_3$ ZRs can be constructed by periodic translation of the unit cells along the direction of $\boldsymbol{a}_1$. The ZR is characterized by the boundary conditions imposed on the wavefunction ${\psi}_s=\begin{pmatrix} \psi_{s}^A & \psi_{s}^B & \psi_{s}^C \end{pmatrix} ^T$ in the real space $\boldsymbol{r}=(x,y)$. ${\psi}_s(\boldsymbol{r})$ is given by a linear combination of the Bloch state $\psi_{s,\boldsymbol{k}}(\boldsymbol{r})=e^{i\boldsymbol{k}\cdot\boldsymbol{r}}| \psi_{s,\boldsymbol{k}}\rangle$. In Fig.~\ref{fig:lett-lattices}(a),  
    \begin{align}
    {\psi}_{s}^{B}(\boldsymbol{O}+l\boldsymbol{a}_1) &= 0,
    \label{eq:l-B}\\
    \alpha\left[ {\psi}_{s}^{C}(\boldsymbol{L}+l\boldsymbol{a}_1) + {\psi}_{s}^{C}(\boldsymbol{L}+(l-1)\boldsymbol{a}_1) \right]+& \nonumber\\{\psi}_{s}^{A}(\boldsymbol{L}-\boldsymbol{b}_2+l\boldsymbol{a}_1)&=0,
    \label{eq:l-AC}
    \end{align}
where $\boldsymbol{O}=(0,0)$ is the origin and $l\in \mathbb{Z}$. Eq.~(\ref{eq:l-B}) describes the missing $B$ sites at $m=0$. Eq.~(\ref{eq:l-AC}) describes the terminations of bonds connecting the $B$ sites at $m=3J $ to the nearest $A$ and $C$ sites outside the ZR. For the bulk states~\cite{delplace2011}, we adopt ${\psi}_s(\boldsymbol{r})=\mathcal{A}\psi_{s,\boldsymbol{k}}(\boldsymbol{r})+\mathcal{A}^\prime\psi_{s,\boldsymbol{k^\prime}}(\boldsymbol{r})$. Eq.~(\ref{eq:l-B}) implies $\mathcal{A}^\prime = - \mathcal{A}$, and $\boldsymbol{k}\cdot\boldsymbol{a}_1=\boldsymbol{k^\prime}\cdot\boldsymbol{a}_1$ due to the conservation of momentum in the $x$ direction. Additionally, the conservation of energy $\varepsilon_{\boldsymbol{k}} = \varepsilon_{\boldsymbol{k^\prime}} $ gives $\boldsymbol{k}\cdot\boldsymbol{a}_1=\boldsymbol{k^\prime}\cdot\boldsymbol{a}_1=(\boldsymbol{k}+\boldsymbol{k^\prime})\cdot\boldsymbol{a}_2$ and therefore $\boldsymbol{k^\prime}=(k_x,-k_y)$.





Now let us define a unitary matrix 
\begin{align}
    M_{\boldsymbol{k}} = \begin{pmatrix}
   e^{i\boldsymbol{k}\cdot(\boldsymbol{b}_2-\boldsymbol{a}_1/2 )} \cos \gamma  & 0 & e^{-i\boldsymbol{k}\cdot\boldsymbol{b}_3 } \sin\gamma \\
   0 & 1 & 0\\
  - e^{i\boldsymbol{k}\cdot(\boldsymbol{b}_2-\boldsymbol{a}_1/2 )} \sin\gamma & 0 &  e^{-i\boldsymbol{k}\cdot\boldsymbol{b}_3 } \cos\gamma
    \end{pmatrix}.
    \label{eq:M}
\end{align}
By setting $\gamma=\tan ^{-1}(\alpha\beta)$ with $\beta=2\cos(k_xa/2)\geq0$ for $k_x\in[-\pi/a,\pi/a]$, $h_{\boldsymbol{k}}$ is transformed into the Hamiltonian of the stub SSH chain in Fig.~\ref{fig:lett-lattices}(b) as follows:
\begin{align}
H_{\boldsymbol{k}} = M_{\boldsymbol{k}}h_{\boldsymbol{k}}M_{\boldsymbol{k}}^\dagger
= -t_M \begin{pmatrix}
     0   & \mathcal{F}_{\boldsymbol{k}}   & 0\\
     \mathcal{F}_{\boldsymbol{k}}^* & 0 & v_C \\
     0 & v_C & 0
\end{pmatrix},
\label{eq:sSSH-Hlett}
\end{align}
where $t_M=t_A\cos\gamma$ and $\mathcal{F}_{\boldsymbol{k}} = v_A + v_A^\prime e^{-i\sqrt{3}k_y a/2 }=|\mathcal{F}_{\boldsymbol{k}}|e^{-i\Phi_{\boldsymbol{k}}}$. In each unit cell (dashed rectangle), the positions of all sites in the $y$ axis are identically given by $Y_j=j\sqrt{3}a/2$. $v_A=\beta(1+\alpha^2)$ [$v_A^\prime=1+\alpha^2\beta^2$] is the intracell [intercell] couplings between the $B$ and $A$ sites. $v_C=\alpha(1-\beta^2)$ is the coupling between the $B$ and $C$ sites. Thus, the unitary transform reduces the $\alpha$-$T_3$ ZR into the stub SSH chain by mapping $k_x$ into the hopping parameters. The eigenvalues of $H_{\boldsymbol{k}}$ are $E_{s,{\boldsymbol{k}}}=st_M\sqrt{|\mathcal{F}_{\boldsymbol{k}}|^2+{v_C}^2}$ and $E_{0,\boldsymbol{k}}=0$. The case $v_C=0$ corresponds to the SSH chain (with additional $E_{0,\boldsymbol{k}}=0$ originating from the uncoupled $C$ sites), where the gap closing at $|k_y|=2\pi/\sqrt{3}a$ indicates a topological phase transition when $v_A/v_A^\prime =1$. On the other hand, by denoting $\Phi_{{\boldsymbol{k}}}$ at $k_y=\pm 2\pi/\sqrt{3}a$ as $\Phi_{\pm}$, it is noted that regardless the value of $v_C$,
\begin{align}
    \Phi_{\pm} =
    \begin{cases}
        \pm\pi~\mathrm{for}~v_A/v_A^\prime<1,\\
        0~\mathrm{for}~v_A/v_A^\prime>1.
    \end{cases}
    \label{eq:lett-Phi}
\end{align}
Eq.~(\ref{eq:lett-Phi}) will be employed to evaluate the $\mathbb{Z}_2$ invariant for the stub SSH chain. By defining $\Theta_{\boldsymbol{k}}=\tan^{-1}(v_C/|\mathcal{F}_{\boldsymbol{k}}|)$, the eigenstates are given by
\begin{align}
       | \Psi_{s,\boldsymbol{k}}\rangle =
  \frac{1}{\sqrt{2}}  \begin{pmatrix}
\cos\Theta_{\boldsymbol{k}} e^{-i\Phi_{\boldsymbol{k}}}\\
s\\
\sin\Theta_{\boldsymbol{k}}
    \end{pmatrix},
    ~| \Psi_{0,\boldsymbol{k}}\rangle =
  \begin{pmatrix}
\sin\Theta_{\boldsymbol{k}} e^{-i\Phi_{\boldsymbol{k}}}\\
0\\
-\cos\Theta_{\boldsymbol{k}}
    \end{pmatrix}.
    \label{eq:psi-chain}
\end{align}

\begin{figure}[t]
\begin{center}
\includegraphics[width=75mm]{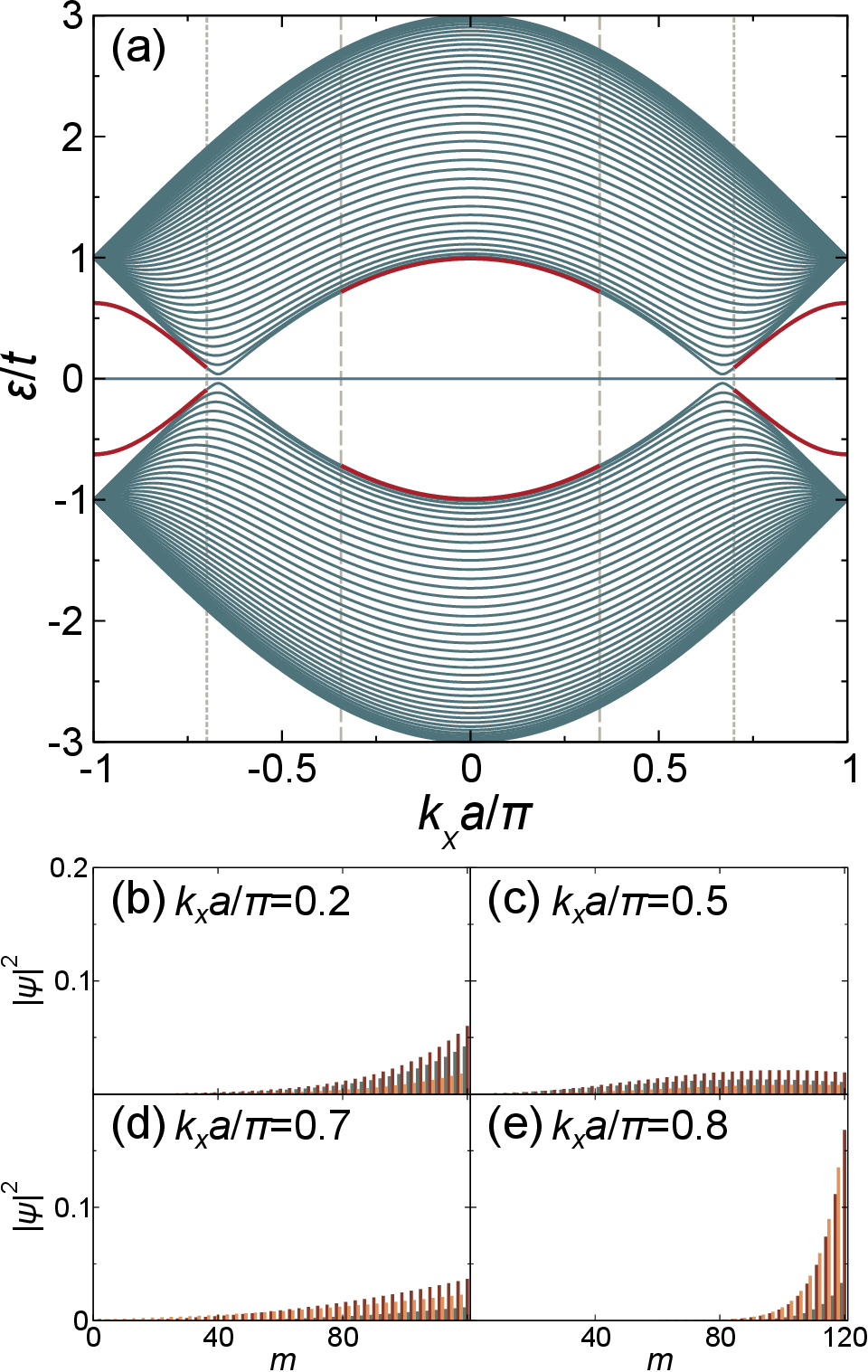}
\caption{ (a) Energy spectra of the $\alpha$-$T_3$ ZR as a function of $k_x$ for $\alpha=0.8$ and $J=40$. The bold red lines indicate the edge states. The transitions from bulk to edge states occur at $|k_x|\approx 0.3434\pi/a$ and $0.6991\pi/a$, accordingly marked by the vertical dashed and dotted lines. Probability density for the in-gap states per site $m$ for (b) $k_x=0.2\pi/a$, (c) $0.5\pi/a$, (d) $0.7\pi/a$, and (e) $0.8\pi/a$. }
\label{fig:lett-E-Psi}
\end{center}
\end{figure}




The wavefunction of the chain is denoted by $\Psi_s=\begin{pmatrix} \Psi_{s}^A & \Psi_{s}^B & \Psi_{s}^C \end{pmatrix} ^T$. As illustrated by Fig.~\ref{fig:lett-lattices}(b), the missing $B$ ($A$) site at $Y_0$ ($Y_{J+1}$) necessitates 
    \begin{align}
    {\Psi}_{s}^{B}(Y_0) &= 0,
    \label{eq:emp-B}\\
    {\Psi}_{s}^{A}(Y_{J+1}) &= 0.
    \label{eq:emp-A}
    \end{align}
The equivalence between the $\alpha$-$T_3$ ZR in Fig.~\ref{fig:lett-lattices}(a) and the stub SSH chain in Fig.~\ref{fig:lett-lattices}(b) is established from the boundary conditions. It can be shown that Eqs. (\ref{eq:emp-B}) and (\ref{eq:emp-A}) are identical to Eqs. (\ref{eq:l-B}) and (\ref{eq:l-AC}), respectively, because $\Psi_s(\boldsymbol{r})=M_{\boldsymbol{k}}\psi_s(\boldsymbol{r})$. Since $\Psi_s(y)=\Psi_{s,\boldsymbol{k}}(y)-\Psi_{s,\boldsymbol{k^\prime}}(y)$, where $\Psi_{s,\boldsymbol{k}}(y)=e^{ik_y y}|\Psi_{s,\boldsymbol{k}}\rangle$, the quantization condition of $k_y$ is derived from Eq.~(\ref{eq:emp-A}) as follows:
\begin{align}
    \sqrt{3}(J+1)k_ya/2-\Phi_{\boldsymbol{k}}=n\pi,~\mathrm{for}~n=1,2,\dots, J. 
    \label{eq:lett-Q2}
\end{align}
Eq. (\ref{eq:lett-Q2}) is solved graphically along $k_y\in(0,2\pi/\sqrt{3}a)$  [$\Psi_s(y)=0$ at $k_y=0$ and $2\pi/\sqrt{3}a$]. It is noted that Eq. (\ref{eq:lett-Q2}) also holds for the SSH chain because $\Phi_{\boldsymbol{k}}$ does not depend on $v_C$. The missing solutions of Eq. (\ref{eq:lett-Q2}) indicate the emergence of edge states~\cite{delplace2011} when
\begin{align}
    v_A/v_A^\prime < 1-1/(J+1).
    \label{eq:lett-Q3}
\end{align}
The ratio $(v_A/v_A^\prime)_J = 1-1/(J+1)$ is the critical value at which the transition of bulk to edge states occur in a finite system~\cite{delplace2011}. On the other hand, the inversion symmetry is broken for $v_C\neq 0$~\cite{bartlett2021illuminating}. Subsequently, the existence of edge states can not be predicted from the Zak phase, which is no longer quantized to $\pi$ or $0$~\cite{zak1989}.


The existence of edge states in the $k_x$ space of the $\alpha$-$T_3$ ZR can be straightforwardly determined from Eq.~(\ref{eq:lett-Q3}). By inserting the definitions of $v_A$ and $v_A^\prime$, we obtain
\begin{align}
    \alpha^2\beta^2 -\frac{1+\alpha^2}{1-(J+1)^{-1}}\beta + 1 > 0.
    \label{eq:lett-quad}
\end{align}
For $\alpha\neq 0$, the solutions of Eq.~(\ref{eq:lett-quad}) are $\beta_J^+<\beta\leq 2$ and $0\leq\beta<\beta_J^- $, where $\beta_J^{\pm}=2\cos(\chi_J^\pm a/2)$. Here, $\chi_J^\pm$ are the critical values of $|k_x|\in[0,\pi/a]$ at which bulk states undergo transitions to edge states appearing in the range
\begin{align}
0\leq|k_x|<\chi_J^+~\mathrm{and}~\chi_J^-<|k_x|\leq{\pi}/{a}.
    \label{eq:range-ZR}
\end{align}
In the limit $J\rightarrow \infty$, $\chi_J^\pm$ converge to 
\begin{align}
    \chi_\infty^+\sim({2}/{a})\cos^{-1}\left({1}/{2\alpha^2}\right),~\chi_\infty^-\sim{2\pi}/{3a}.
    \label{eq:cr-ZR}
\end{align}
Since $\chi_\infty^+$ is undefined for $\alpha\in(0, 1/\sqrt{2})$, the range of existence for the edge states is identical to that of the graphene ZR~\cite{wakabayashi2010electronic,delplace2011}. 
The edge states in the $T_3$ ZR ($\alpha=1$) exist at all $|k_x|$ excluding $\chi_\infty^+=\chi_\infty^-=2\pi/3a$.



The energy of the edge states $\tilde{\varepsilon}_s$ is deduced as follows. In the SSH chain, $\tilde{E}_s\sim 0$ due to the chiral symmetry~\cite{ryu2002,delplace2011}. Accordingly for $v_C\neq 0$, $\tilde{E}_s \sim st_M|v_C|$, where the state $s=-1$ ($+1$) is analogous to the bonding (antibonding) molecular orbital. Since the eigenvalues are invariant under the unitary transform, $\tilde{\varepsilon}_s=\tilde{E}_s=st_C\cos\gamma|1-\beta^2|$. Thus, the edge states in the $\alpha$-$T_3$ ZR are dispersive due to the dependence on $\beta$. At $|k_x|\approx 2\pi/3a$ and $|k_x|=\pi/a$, we get $\tilde{\varepsilon}_s\approx 0$ and $\tilde{\varepsilon}_s= st_C$, respectively.

Fig.~\ref{fig:lett-E-Psi}(a) shows the energy spectra as a function of $k_x$ for $\alpha=0.8$ and $J=40$. In addition to the dispersive bands, the spectra include a $J$-fold degenerate flat band at $\varepsilon=0$. The vertical dashed and dotted lines are located at $|k_x|=\chi_{40}^+\approx0.3434\pi/a$ and $\chi_{40}^-\approx 0.6991\pi/a$, respectively. The bold red lines depict the edge states. Figs.~\ref{fig:lett-E-Psi}(b)-(e) show the probability density $|\psi|^2$ of the in-gap states at the site $m$ at several $k_x$'s. Figs.~\ref{fig:lett-E-Psi}(b) and (e) clearly show the spatial distribution of asymmetric edge states, where $|\psi|^2$'s are concentrated around the top edge of the ZR and exponentially decay towards the bottom edge. In Fig.~\ref{fig:lett-E-Psi}(d), the decay of $|\psi|^2$ is not pronounced because the value of $k_x$ is close to $\chi_{40}^-$. Fig.~\ref{fig:lett-E-Psi}(c) exhibits the profile of bulk states where $|\psi|^2$ is extended throughout the ZR.

\begin{figure*}[t]
\begin{center}
\includegraphics[width=174mm]{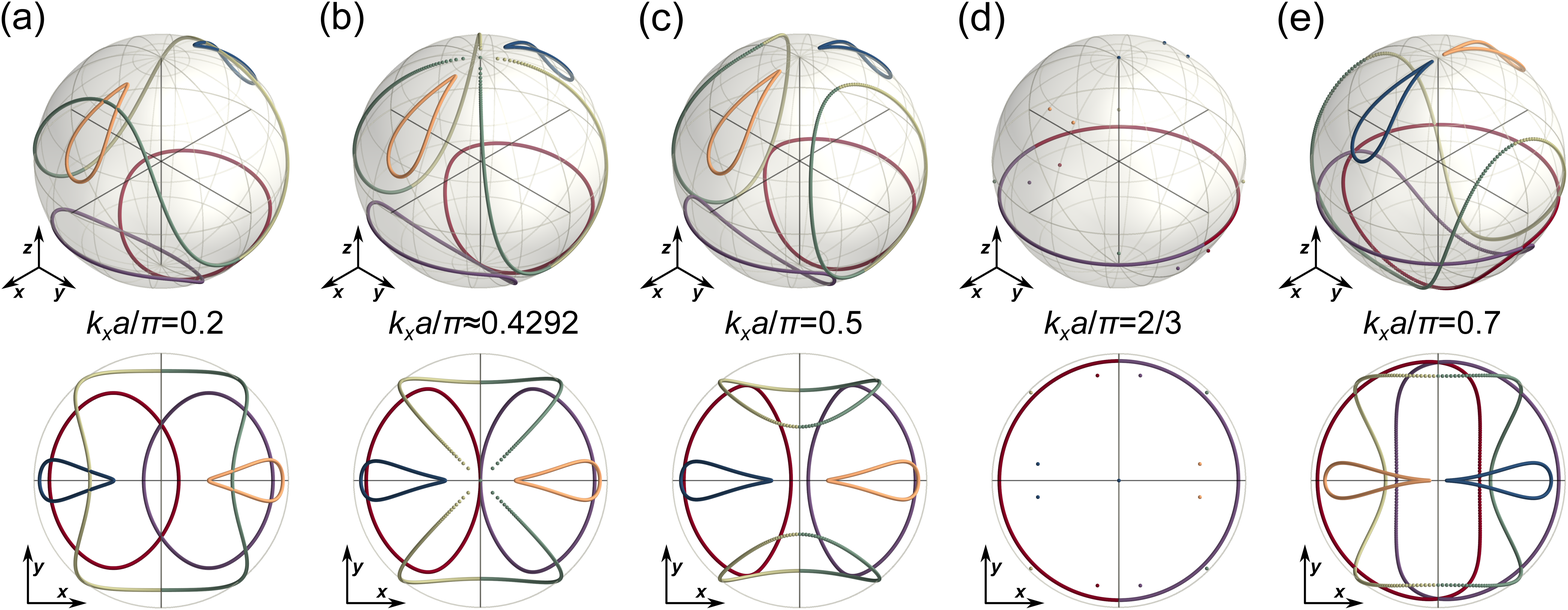}
\caption{ Trajectories of the Majorana stars on the Bloch sphere (top panels) and the $xy$ plane (bottom panels) for $\alpha=0.8$, at (a) $k_x=0.2\pi/a$, (b) $\chi_\infty^+\approx 0.4292\pi/a$, (c) $0.5\pi/a$, (d) $\chi_\infty^-= 2\pi/3a$, (e) $0.7\pi/a$. The red and yellow [blue and purple] curves correspond to the valence [conduction] band. The light- and dark-green curves correspond to the flat band. }
\label{fig:lett-Maj-C2}
\end{center}
\end{figure*}

The edge states are topologically characterized using the Majorana representation~\cite{majorana1932atomi,liu2014representation}. By treating the eigenstate $|\Psi\rangle$ of an $N\times N$ Hamiltonian as a pseudospinor for $S=(N-1)/2$, $|\Psi\rangle$ can be represented by a set of the Pauli pseudospinors $|\zeta_\mu\rangle = \begin{bmatrix} \cos(\eta_\mu/2) & \sin(\eta_\mu/2)e^{i\xi_\mu}  \end{bmatrix}^{T}$, where $\eta$ and $\xi$ are the polar and azimuthal angles of the Bloch sphere. The trajectory of each Majorana 'star' $\boldsymbol{\zeta}_\mu = \langle \zeta_\mu | \boldsymbol{\sigma} | \zeta_\mu \rangle$ thus visualizes the evolution of $|\Psi\rangle$.

In the Schwinger theory of angular momentum~\cite{schwinger1952},  $|\Psi\rangle$ can be constructed by the bosonic creation operators $\hat{a}_{\uparrow,\downarrow}^{\dagger}$ acting on the vacuum state $|\varnothing \rangle$ as follows:
\begin{align}
    |\Psi \rangle  = \frac{\mathcal{C}_S}{\sqrt{(2S)!}}\prod_{\mu=1}^{2S}\left[  \hat{a}_{\uparrow}^{\dagger} + \lambda_\mu  \hat{a}_{\downarrow}^{\dagger}         \right ]| \varnothing  \rangle,
    \label{eq:lett-Maj1}
\end{align}
where $\mathcal{C}_S$ is the amplitude of $|\Psi \rangle$, and $\lambda_\mu=\tan (\eta_\mu/2)e^{i\xi_\mu}$ is obtained by solving the Majorana polynomial~\cite{majorana1932atomi} 
\begin{align}
    \sum_{\mu=0}^{2S}(-1)^\mu\frac{ \mathcal{C}_S-\mu}{\sqrt{(2S-\mu)!\mu !}}\lambda^{2S-\mu} = 0.
    \label{eq:lett-Maj2}
\end{align}
In particular for the stub SSH chain, $S=1$ and $|\Psi\rangle = \begin{pmatrix}\mathcal{C}_1 & \mathcal{C}_0 & \mathcal{C}_{-1}\end{pmatrix}^{T}$ is given by Eq. (\ref{eq:psi-chain}). For $c=s$ and $c=0$, Eq.~(\ref{eq:lett-Maj2}) accordingly reduces to
\begin{subequations}
     \begin{align}
    \cos\Theta_{\boldsymbol{k}}e^{-i\Phi_{\boldsymbol{k}}}{\lambda^2} -\sqrt{2}s\lambda+{\sin\Theta_{\boldsymbol{k}}} = 0,\label{eq:polys}\\
        {\sin\Theta_{\boldsymbol{k}}}e^{-i\Phi_{\boldsymbol{k}}}{\lambda^2} -{\cos\Theta_{\boldsymbol{k}}} = 0.
    \label{eq:poly0}
\end{align}
\end{subequations}
Let $\lambda_{s,\boldsymbol{k}}^\mp$ and $\lambda_{0,\boldsymbol{k}}^\mp$ be the roots of Eqs.~(\ref{eq:polys}) and (\ref{eq:poly0}), respectively. From the coefficients of the quadratic equations, it is clear that $\Lambda_{s,\boldsymbol{k}}=\prod_{\nu=\mp}\lambda_{s,\boldsymbol{k}}^\nu=\tan\Theta_{\boldsymbol{k}}e^{i\Phi_{\boldsymbol{k}}}$, and $\Lambda_{0,\boldsymbol{k}}=\prod_{\nu=\mp}\lambda_{0,\boldsymbol{k}}^\nu=-\mathrm{cot}\Theta_{\boldsymbol{k}}e^{i\Phi_{\boldsymbol{k}}}$.

The $\mathbb{Z}_2$ topological index is defined by the azimuthal winding number $W_c$, which counts for the total number of times $\boldsymbol{\zeta}_{c,\boldsymbol{k}}^\nu$'s travel around the $z$ axis of the Bloch sphere. $W_c$ is calculated as follows~\cite{bartlett2021illuminating}:
\begin{align}
    {W}_c = \frac{1}{2\pi}   \int_\mathrm{BZ} dk_y\frac{\partial}{\partial k_y} \sum_{\nu=\mp}\xi_{c,\boldsymbol{k}}^{\nu}.
     \label{eq:lett-Wc1}
\end{align}
Since $\lambda_{c,\boldsymbol{k}}^{\nu} = \tan(\eta_{c,\boldsymbol{k}}^{\nu}/2)\mathrm{exp}({i\xi_{c,\boldsymbol{k}}^{\nu}})$ by definition, therefore $\sum_{\nu=\mp}\xi_{c,\boldsymbol{k}}^{\nu}=\Phi_{\boldsymbol{k}}$. By inserting Eq.~(\ref{eq:lett-Phi}) into Eq.~(\ref{eq:lett-Wc1}), 
\begin{align}
 {W}_{c} =
 \begin{cases}
     1~\mathrm{for}~ v_A/v_A^\prime<1,\\
    0~\mathrm{for}~ v_A/v_A^\prime>1.
 \end{cases}
\label{eq:lett-Wc2}
 \end{align}
As $k_y$ traverses across the 1D BZ, the trajectories of $\boldsymbol{\zeta}_{c,\boldsymbol{k}}^\nu$'s enclose the $z$ axis at $k_x$'s that carry the edge states.


Figs.~\ref{fig:lett-Maj-C2}(a)-(e) show the trajectories of $\boldsymbol{\zeta}_{c,\boldsymbol{k}}^\nu$'s on the Bloch sphere (top panels) and the $xy$ plane (bottom panels) for $\alpha=0.8$ at (a) $k_x=0.2\pi/a$, (b) $\chi_\infty^+\approx 0.4292\pi/a$, (c) $0.5\pi/a$, (d) $\chi_\infty^-=2\pi/3a$, and (e) $0.7\pi/a$. The red/yellow, light-green/dark-green, and blue/purple curves correspond to $c=-1$, $0$, and $+1$, respectively. In Figs.~\ref{fig:lett-Maj-C2}(a) and (e), the red, purple, and combined green loops wind around the $z$ axis once, thus $W_c=1$. Conversely, in Fig.~\ref{fig:lett-Maj-C2}(c), none of the loops enclose the $z$ axis, so $W_c=0$. At $\chi_\infty^\pm$, $W_c$ becomes ill-defined for different reasons. In Fig.~\ref{fig:lett-Maj-C2}(b), the red, purple, and combined green curves intersect the poles. This behavior indicates a topological phase transition without gap closing~\cite{ezawa2013topological}. In Fig.~\ref{fig:lett-Maj-C2}(d), the trajectories of $\zeta_c^{\nu}$'s are discontinuous due to the gap closing at $|k_y|= 2\pi/\sqrt{3}a$. Here, the system is reduced to the metallic SSH chain because $\beta=1$ and consequently $v_C=0$, $v_A=v_{A1}^\prime$. In the case of $T_3$ ZR, $W_c=1$ for all except at $|k_x|=2\pi/3a$. Thus, the gap closing in the corresponding stub SSH chain is not accompanied by a topological phase transition.

Previously, Liu and Wakabayashi~\cite{liu2017novel} proposed a 2D SSH model~\cite{liu2017novel,obana2019} exhibiting a topologically non-trivial phase with zero Berry curvature, due to the conserved TRS and inversion symmetry. The $\mathbb{Z}_2$ invariant is given by 2D Zak phase calculated by the integration of the Berry connection over the BZ. In this study, we identify $T_3$ ZR as a topologically non-trivial system with zero Berry phase, where the $\mathbb{Z}_2$ invariant is defined by $W_c$.

\begin{figure}[t]
\begin{center}
\includegraphics[width=60mm]{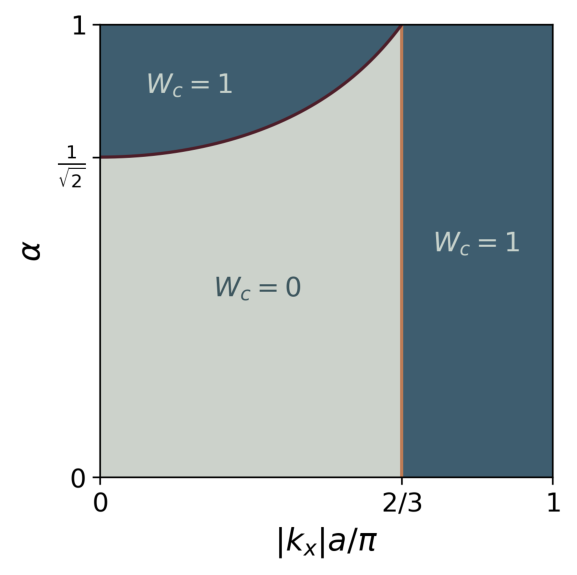}
\caption{ Topological phase diagram of the $\alpha$-$T_3$ ZR. The dark
(light) region indicates the non-trivial (trivial) phase, where the edge states are present (absent). The brown (orange) line marks the phase transition without (with) the gap closing.  }
\label{fig:lett-phase}
\end{center}
\end{figure}

Fig.~{\ref{fig:lett-phase}} shows the topological phase diagram of the $\alpha$-$T_3$ ZR. The dark (light) region indicates the non-trivial (trivial) phase, where the presence (absence) of edge states are characterized by $W_c=1$ ($W_c=0$). The topological phase transition with (without) gap closing is marked by the brown (orange) line.

Since the Hamiltonians of the $\alpha$-$T_3$ lattice and stub SSH chain are equivalent, the bulk topological property is conserved when the dimension is compacted. Recently, C\'{a}ceres \textit{et al.}~\cite{caceres2022experimental} realized the stub SSH chain in an experiment using a photonic lattice. They demonstrated that for $v_A/v_A^\prime<1$, the edge states appear at $E_s=st|v_c|$ (consistent with the present result), and persist against disorders that perturb $v_A$ and $v_A^\prime$. We analytically prove that the edge states are characterized by $W_c$, which may explain the topological origin of the robustness. 

In summary, we formulate the bulk-boundary correspondence for the massless Dirac fermion with a non-quantized Berry phase in the $\alpha$-$T_3$ lattice, which is mapped into the stub SSH chain by the unitary transform of the Hamiltonian. In the Majorana representation of the eigenstates, the $\mathbb{Z}_2$ topological index is defined from the azimuthal winding number on the Bloch sphere. Our method also shows that the $T_3$ lattice is topologically non-trivial despite the zero Berry phase. The existence of edge states in distinct $\alpha$-$T_3$ ribbon configurations are discussed in detail in the accompanying paper~\cite{pratama24-main}. 

We thank R. Saito, N. T. Hung, S. Hayashi, and E. H. Hasdeo for helpful discussions. This work is partly supported by Grant-in-Aid for Scientific Research, JSPS KAKENHI (23K03293), and JST-CREST (JPMJCR18T1).

\bibliographystyle{pratrev}
\bibliography{pratama}

\end{document}